\documentclass[pra,reprint,amsfonts,amssymb,showpacs]{revtex4-1}
\usepackage{graphicx}
\usepackage{bm}

\begin{document}

\title{
Design method of dynamical decoupling sequences integrated with optimal control theory}
\author{Yutaka Tabuchi}
\email{tabuchi@qc.ee.es.osaka-u.ac.jp}
\author{Masahiro Kitagawa}
\affiliation{Graduate School of Engineering Science, Osaka University, Toyonaka, Osaka 560-8531, Japan}

\date{\today}

\begin{abstract}
A method for synthesizing dynamical decoupling (DD) sequences is presented, which can tailor these sequences to a given set of qubits, environments, instruments, and available resources using partial information of the system. The key concept behind the generation of the DD sequences involves not only extricating the strong dependence on the coupling strengths according to the ``optimal control,'' but also exploiting the ``refocus'' technique used conventionally to obtain DD sequences. The concept is a generalized one that integrates optimal control and designing of DD sequences. 
\end{abstract}


{\maketitle}

\section{Introduction}
In quantum information processing (QIP), dynamical decoupling (DD)~\cite{bib:Zanardi99,bib:Viola99} is a useful tool for suppressing the decoherence resulting from multiple couplings among qubits and their environments. Potential applications of dynamical decoupling to quantum memory~\cite{bib:Khodjasteh05}, quantum computation (QC)~\cite{bib:Viola99_2,bib:Byrd03,bib:West10_2,bib:Khodjasteh10}, and fault-tolerant quantum computation~\cite{bib:Ng11} have been proposed recently. 
Trains of successive pulses such as concatenated DD, Uhrig DD~(UDD), and quadratic DD~(QDD)~\cite{bib:Khodjasteh05,bib:Uhrig07,bib:West10} have been proposed and demonstrated successfully to eliminate unwanted couplings~\cite{bib:Biercuk09,bib:Peng11,bib:Bylander11} without knowledge of surrounding environment, e.g., its coupling strengths or structures. These methods are useful for experiments in which system details are largely unavailable. 
In practical experiments, however, theoretical approximations such as the delta pulse approximation result in the \textit{finite pulse width} problem~\cite{bib:Khodjasteh07}, and the requirement of an infinite frequency bandwidth for a rectangular pulse causes deformation at the leading and trailing edges of the pulse~\cite{bib:Barbara91,bib:tabuchi10}. Both these practical resource limitations cause serious bottlenecks when using pulse DD sequences. 
In contrast, numerically optimized pulse sequences~\cite{bib:Peirce88,bib:Brion11,bib:Khaneja05} make use of the complete qubit- and environment-related information. However, these methods are attractive~\cite{bib:Zhang11} only when one can obtain details of a qubit-bath system, although spectroscopy and generation of pulse sequences become difficult as the system size increases. 
Hence, when using partial system information, we can expect better performance from DD sequences. Previous studies that have used this approach include refs.~\cite{bib:Gordon08,bib:Clausen10}, in which the spectra of the surrounding environment were reflected on their energy-constrained decoupling waveforms.

In this paper, we propose a method to numerically synthesize DD sequences under realistic resource limitations such that they can be tailored to a system based on available knowledge about qubits, environments, and instruments. In the two examples given in this paper, we generate bandwidth- and energy-constrained DD waveforms that are free from the finite pulse width problem and are robust to waveform deformation arising from instruments~\cite{bib:tabuchi10}. The paper is organized as follows: In Sec.~\ref{sec:basic}, we describe the basic concept of our DD design. In Sec.~\ref{sec:dephase}, we present a pragmatic example in which a single qubit is coupled to many two-level systems, and where the control field has imperfections due to instruments. In Sec.~\ref{sec:mrev}, we apply our method to a multiqubit system where the qubits couple to each other. In Sec.~\ref{sec:resource}, we discuss the performance of DD sequences generated under resource restrictions in terms of energy and bandwidth. The appendices describe how the average Hamiltonians are calculated using Floquet Hamiltonian theory in the examples.

\section{Basic Concept\label{sec:basic}}
Let $\hat{\mathcal{H}}_0$ be the Hamiltonian of a system under consideration, and $\hat{V}(t)$ be a time-dependent external perturbation, represented in an appropriate rotating frame. We use $\hbar = 1$ units throughout. The system Hamiltonian $\hat{\mathcal{H}}_0$ can be expressed as
\begin{equation}
\hat{\mathcal{H}}_0 = \sum_\alpha h_\alpha \hat{\Theta}_\alpha,
\label{eq:systemh}
\end{equation} using a set of orthonormal Hermitian operators $\{\hat{\Theta}_\alpha | \hat{\Theta}_\alpha = \hat{\Theta}_\alpha^\dagger, \text{Tr}[\hat{\Theta}_\alpha\hat{\Theta}_\beta]=\delta_{\alpha\beta}\}$ and their coefficients $\{h_\alpha\}$. 
For example, a system Hamiltonian that describes the coupling between a single qubits and the surrounding environment can be expanded with operators $\hat{\Theta}_\alpha = \frac{1}{2}\hat{\sigma}_\alpha \otimes \hat{B}_\alpha$ ($\alpha = x, y, z$), where $\hat{\sigma}_\alpha$ are the qubit's Pauli operators and $\hat{B}_\alpha$ are bath operators. 
%
%$\hat{V}(t) = \sum_\alpha v_\alpha(t) \hat{\Theta}_\alpha$ 
%
Consider $\hat{\mathcal{H}}_0$ in the interaction frame of $\hat{V}(t)$ using unitary operator $\hat{U}_V(t) = \hat{\mathcal{T}}\exp(i\int_0^t\hat{V}(t^\prime)\,dt^\prime)$, where $i = \sqrt{-1}$ and $\hat{\mathcal{T}}$ is the time-ordering operator. The Hamiltonian of Eq.~(\ref{eq:systemh}) in the frame of $\hat{V}(t)$ then becomes the \textit{modulated} Hamiltonian $\hat{\widetilde{\mathcal{H}}}_{0}(t)$ as follows:
\begin{equation}
 \hat{\widetilde{\mathcal{H}}}_{0}(t) = \hat{U}_V(t) \hat{\mathcal{H}}_0 \hat{U}_V^\dagger (t)
  = \sum_\alpha h_\alpha \left( \sum_\beta c_{\beta \alpha}(t) \hat{\Theta}_\beta \right),
\end{equation}
where $\hat{U}_V(t) \hat{\Theta}_\alpha \hat{U}_V^\dagger (t) = \sum_\beta c_{\beta \alpha}(t) \hat{\Theta}_\beta$. In this process, the effect of finite pulse width is naturally incorporated into $c_{\beta \alpha}(t)$. Delta pulse approximation of $\hat{V}(t)$ can be also treated as discontinuities of $c_{\beta\alpha}(t)$. We call $c_{\beta\alpha}(t)$ a system-modulation matrix, as used in ref.~\cite{bib:Clausen10}.

This system can be stroboscopically ``time-suspended,'' that is, decoupled from environment, if there is an external perturbation $\hat{V}(t)$ such that the average Hamiltonian $\hat{\overline{\mathcal{H}}}_{\text{eff}}$ of $\hat{\widetilde{\mathcal{H}}}_0(t)$ vanishes~\cite{bib:Haeberlen68,bib:Viola99,bib:Brion11}, and if the period $T$ of $\hat{\overline{\mathcal{H}}}_{\text{eff}}$ satisfies $\|\hat{\mathcal{H}}_0T\| < 1$ so that Magnus expansion of the evolution operator converges. 
Suppose $\hat{\widetilde{\mathcal{H}}}_0(t)$ has $T$-periodicity; then, the average Hamiltonians~\cite{bib:Haeberlen68,bib:Viola99} up to the first order become
\begin{widetext}
\begin{eqnarray}
 \hat{\overline{\mathcal{H}}}^{(0)}_{\text{eff}}
  &=& \frac{1}{T} \int_0^T \hat{\widetilde{\mathcal{H}}}_0(t^\prime)\,dt^\prime 
  = \sum_{\alpha,\beta} h_\alpha \left(\frac{1}{T}\int_0^T c_{\beta\alpha}(t^\prime) dt^\prime \right)\hat{\Theta_\beta}, \label{eq:ave0}\\
 \hat{\overline{\mathcal{H}}}^{(1)}_{\text{eff}}
  &=& \frac{1}{2iT} \sum_{\alpha,\alpha^\prime} h_\alpha h_{\alpha^\prime}
         \sum_{\beta < \beta^\prime} 
          \left(
           \int_0^T\!dt_2 \int_0^{t_2}\!dt_1\,\,
           \{c_{\beta^\prime \alpha^\prime}(t_2) c_{\beta \alpha}(t_1) 
          - c_{\beta^\prime \alpha}(t_1) c_{\beta \alpha^\prime}(t_2)\}
         \right) [\hat{\Theta}_{\beta^\prime},\hat{\Theta}_{\beta}]. 
         \label{eq:ave1} 
\end{eqnarray}
\end{widetext}
We can obtain DD sequences $\hat{V}(t)$ by minimizing the coefficients of the operators $\hat{\Theta}_\beta$ and $[\hat{\Theta}_{\beta^\prime},\hat{\Theta}_\beta]$, i.e., the terms within the large curly brackets in Eqs.~(\ref{eq:ave0}) and~(\ref{eq:ave1}).
Given that the external perturbation $\hat{V}(t)$ does not always guarantee the $T$-periodicity of $\hat{\widetilde{\mathcal{H}}}_0(t)$, the system-modulation  matrix $c_{\beta\alpha}(t)$ should be constrained such that $c_{\beta\alpha}(t)$ are continuous at $t=0,T,\cdots$. 

For numerical minimization, $\hat{V}(t)$ should be parameterized with pulse timings, flip angles, phases, and so on. When the parameters are denoted by $\bm{\zeta} = \{\zeta_1, \zeta_2, \cdots \}$, the system-modulation matrix is also parameterized such that $c_{\beta\alpha}(t) = c_{\beta\alpha}(t;\bm{\zeta})$. A cost function for the numerical minimization is constructed such that 
\begin{equation}
\Phi(\bm{\zeta}) = \sum_{\alpha, \beta}\Phi_{\alpha\beta}^{(0)}(\bm{\zeta}) + \sum_{\alpha^\prime\alpha,\beta^\prime < \beta}\Phi_{\alpha^\prime\alpha,\beta^\prime\beta}^{(1)}(\bm{\zeta}) + \cdots,
\end{equation}
 together with a penalty function of the continuity of $c_{\beta\alpha}(t)$ at $t=0,T$,
\begin{equation}
\Phi_P(\bm{\zeta}) = \sum_{\alpha,\beta} |c_{\beta\alpha}(T;\bm{\zeta})-c_{\beta\alpha}(0;\bm{\zeta})|^2, 
\end{equation}
where 
\begin{eqnarray}
\Phi_{\alpha\beta}^{(0)}(\bm{\zeta}) &=& T^{-1}\biggl| \int_0^T c_{\beta\alpha}(t^\prime;\bm{\zeta}) dt^\prime \biggr|^2 
  \label{eq:cost0} \\
\Phi_{\alpha^\prime\alpha,\beta^\prime\beta}^{(1)}(\bm{\zeta}) &=& T^{-1}\biggl|\int_0^T\,dt_2 \int_0^{t_2}\,dt_1 \nonumber \\
& & \times \left\{c_{\beta^\prime\alpha^\prime}(t_2;\bm{\zeta}) c_{\beta\alpha}(t_1;\bm{\zeta}) \right. \nonumber \\
& & \left. - c_{\beta^\prime\alpha}(t_1;\bm{\zeta}) c_{\beta\alpha^\prime}(t_2;\bm{\zeta})\right\}\biggr|^2. 
  \label{eq:cost1}
\end{eqnarray}
Consideration of higher-order perturbations would result in better decoupling performance; however, doing so would involve cumbersome hand calculations for deriving average Hamiltonians. 

The synthesis of DD sequences is then reduced to a nonlinear minimization problem: 
\begin{eqnarray}
  \min_{\bm{\zeta}}\Phi(\bm{\zeta})\,\,
  \text{subjected to}\,\,\Phi_P(\bm{\zeta}) = 0, 
\end{eqnarray}
which can be numerically performed on a modern digital computer.

This method is basically the same as that given by Eq.~(10) in \cite{bib:Brion11} except that the cost function is constructed in Lie algebra. However, this difference allows us to choose among sequences of DD type, optimal control type~\cite{bib:Peirce88,bib:Khaneja05,bib:Brion11}, or a combination of both these types. 
In general, the expansion coefficients $\{h_\alpha\}$ in Eq.~(\ref{eq:systemh}) contain the details of the system, e.g., coupling strengths and frequency shifts, which can be obtained through spectroscopy. 
In the synthesis of DD type sequences, which do not require knowledge of the qubit's surroundings, $\{h_\alpha\}$ should be excluded from the cost functions in Eqs.~(\ref{eq:cost0}) and (\ref{eq:cost1}) so that the generated sequences do not depend on them. 
For the case where $\{h_\alpha\}$ are known, sequences can be synthesized with their help. If we explicitly include $\{h_\alpha\}$ in a cost function, for example, Eq.~(\ref{eq:cost0}) becomes
\begin{equation}
\Phi_{\beta}^{(0)}(\bm{\zeta}) = T^{-1}\biggl| 
  \sum_\alpha h_\alpha \int_0^T c_{\beta\alpha}(t^\prime;\bm{\zeta}) dt^\prime \biggr|^2, 
\end{equation}
and the obtained sequences suppress decoherence using that information. In the limit of $\|\hat{\mathcal{H}}T\|\rightarrow 0$, or in the limit of $n_0 \rightarrow 0$, where $n_0$ is the maximum order of an average Hamiltonian, this method corresponds to unitary matrix based optimal control. 
When $\{h_\alpha\}$ are partially known, they can be incorporated into the cost function to produce a hybrid type of DD and optimal control. This would be advantageous when we know the qubit-qubit interaction strengths that are to be used for gate operations, but do not know the strengths of qubit-environment couplings that cause decoherence. 

Furthermore, cost functions in Lie algebra alleviate the growth of computational complexity as a system becomes larger. Given that most operators in Lie algebra, which stand for many-body interactions, barely appear, there is a dramatic decrease in computational costs associated with numerical optimization for a system Hamiltonian that is simple and highly symmetric, i.e., it is described with a small number of operators in Lie algebra.
We give a paradigmatic explanation in which a Hamiltonian $\hat{\mathcal{H}}_0 = \sum_{k=0}^n g^{(k)}\hat{\sigma}_{z,k}$ dictates a system, where $\sigma_{z,k}$ is a $k$th qubit's Pauli  $z$-operator, $\{g^{(k)}\}$ are inhomogeneously broadened energy shifts, and $n$ is the number of qubits. In optimal control type, to obtain a DD sequence (time-suspension sequence), we have to compute $2^n \times 2^n$ matrices, which are intractable for large $n$. However, a well-known method that flips the sign of $\hat{\sigma}_{z,k}$ in the frame of $\hat{V}(t)$ by applying consecutive $\pi$ pulses to the qubits, such as CPMG~\cite{bib:Meiboom58}, refocuses the inhomogeneity. Our method automates such a flipping procedure for a given system.

Note that while synthesizing DD sequences, $\hat{\mathcal{H}}_0$ should be \textit{modulated} by $\hat{U}_V(t)$. This requirement is equivalent to the bracket generation condition mentioned in ref.~\cite{bib:Brion11}.

In the following two examples, in order to impose a frequency bandwidth limitation on $\hat{V}(t) = \sum_\alpha v_\alpha(t)\hat{\Theta}_\alpha$, we use Fourier coefficients of \begin{equation}v_\alpha(t) = \sum_{n=1}^p v_{\alpha,n} \sin(2n\pi t/T) \label{eq:waveff} \end{equation} as the parameters of $\hat{V}(t)$, so that $\bm{\zeta}=\{v_{\alpha,n}\}$.
 The Floquet average Hamiltonian theory~\cite{bib:Scholz10} is useful to calculate the average Hamiltonian when a system-modulation matrix is $T$-periodic and expressed as the Fourier expansion $c_{\beta\alpha}(t)=\sum_n c_{\beta\alpha,n} \exp{(2i\pi nt/T)}$. A brief explanation of this calculation is given in Appendix~\ref{sec:floquet}, and the cost functions in the following examples are built using the theory.

\section{Dephasing Problem\label{sec:dephase}}
Let us consider the qubit dephasing problem. A single qubit is coupled to a bath consisting of many two-level systems (TLSs) without energy relaxation. When there exist inevitable pulse errors caused by instruments, an experimenter might need a DD sequence that is robust to these errors. In this example, we show how to incorporate Tycko's composite pulses~\cite{bib:Tycko83} into DD designing. In ref.~\cite{bib:Tycko83}, a Hamiltonian of an external perturbation to control the qubits is described as $\hat{V}(t)+\hat{\mathcal{H}}_{E}(t)$, where $\hat{V}(t)$ denotes an ideal operation to the qubit and $\hat{\mathcal{H}}_{E}(t)$ represents systematic errors due to the instruments. Self-compensating composite pulses are built so that the ideal pulsing of $\hat{V}(t)$ decouples error Hamiltonian $\hat{\mathcal{H}}_{E}(t)$. 

\begin{figure}
 \centering
 \includegraphics{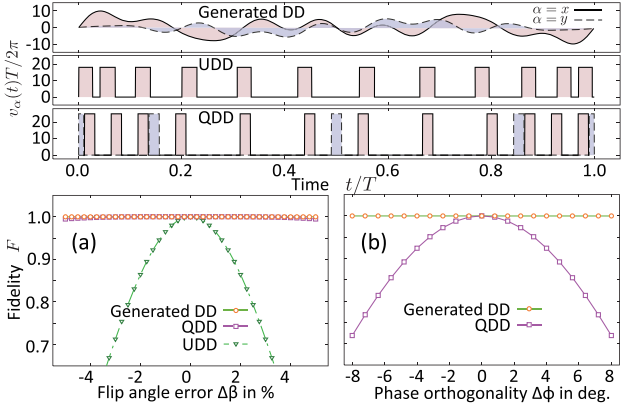}
 \caption{\label{fig:error}(color online). Decoupling fidelities against (a) flip angle error and (b) phase orthogonality error for a single qubit coupled to four bath TLSs with UDD$_{12}$ and QDD$_{3}$ sequences, and the sequence generated in this example. The waveforms of these sequences are displayed in the top panel of the figure.} 
\end{figure}

Let the Hamiltonian of the system be $\hat{\mathcal{H}}_0 = \hat{\mathcal{H}}_{Q-B} + \hat{\mathcal{H}}_{E}(t)$ and the ideal external perturbation be $\hat{V}(t) = v_x(t)\hat{\sigma}_x + v_y(t)\hat{\sigma}_y$, where the qubit-bath interaction is $\hat{\mathcal{H}}_{Q-B} = \sum_k g^{(k)} \hat{\sigma}_z \hat{\sigma}_{z,k}$ and the pulse error term is $\hat{\mathcal{H}}_{E} = \varepsilon_1 v_x(t) \hat{\sigma}_x + \varepsilon_3 v_y(t) \hat{\sigma}_y + \varepsilon_2 v_y(t)\hat{\sigma}_x + \varepsilon_4 v_x(t)\hat{\sigma}_y$. Here, $\hat{\sigma}_\alpha$ and $\hat{\sigma}_{\alpha,k}$ denote Pauli operators of the qubit and the $k$th TLS, respectively. $\varepsilon_1$ and $\varepsilon_3$ denote the error amplitudes of flip angle error, while $\varepsilon_2$ and $\varepsilon_4$ denote the error amplitudes of phase orthogonality error arising, for example, from microwave IQ mixers or hybrid circuits.
The external perturbation $\hat{V}(t)$ modulates the Hamiltonian $\hat{\mathcal{H}}_0$ in its interaction frame, so that 
\begin{eqnarray}
\hat{\widetilde{\mathcal{H}}}_0(t) &=& \sum_k\sum_\beta g^{(k)} c_{\beta z}(t) \hat{\sigma}_{\beta}\hat{\sigma}_{z,k} \nonumber \\
& & + \sum_\beta\{\varepsilon_1v_x(t)c_{\beta x}(t) + \varepsilon_3v_y(t)c_{\beta y}(t) \nonumber \\
& & + \varepsilon_2v_y(t)c_{\beta x}(t) + \varepsilon_4v_x(t)c_{\beta y}(t)\} 
\hat{\sigma}_{\beta}\nonumber,
\end{eqnarray}
where $\sum_\beta c_{\beta\alpha}(t) \hat{\sigma}_{\beta} = \hat{U}_V(t) \hat{\sigma}_\alpha \hat{U}_V^\dagger (t)$. In this example, the zeroth-order cost function is defined as 
\begin{eqnarray}
\Phi^{(0)} &=& \sum_\beta\left|\int_0^T c_{\beta z}(t) dt\right|^2
 \! +\sum_\beta\left|\int_0^T v_x(t)c_{\beta x}(t) dt \right|^2 \nonumber 
 + \cdots, 
\end{eqnarray}
and the first-order cost function is also defined in the same manner~(See Appendices~\ref{sec:ex1_ah} and \ref{sec:ex1_cost} for detailed derivations). Exclusion of $g^{(k)}$ and $\varepsilon_j$ ($j = 1, \cdots, 4$) from the overall cost function gives DD type pulse sequences with a tolerance to pulse imperfections.
DD sequences generated with this method can be tailored to the system by multiplying deliberate weights to terms in the cost function. We minimized the cost function by optimizing $\bm{\zeta} = \{v_{\alpha,n}\}$ using a steepest descent method in combination with a genetic algorithm. The generated waveform expressed in Fourier series form, $v_\alpha(t) = \sum_{n=1}^{9} v_{\alpha,n} \sin(2\pi nt/T)$, as listed in Table~\ref{tab:ex1}, is shown in the top panel of Fig.~\ref{fig:error}.
\begin{table}
  \caption{\label{tab:ex1}Parameters of the DD sequence generated in Sec.~\ref{sec:dephase}}
  \begin{tabular}{c|rlr}
    $n$ & $v_{x,n}T/\pi$ & \hspace{1em} & $v_{y,n}T/\pi$ \\
  \hline
    1 & -0.7030256 & & -3.6201768 \\
    2 &  3.3281747 & & 3.8753985 \\
    3 &  11.390077 & & -1.2311919 \\
    4 &  2.9375301 & & -0.2998110 \\
    5 & -1.8758792 & & 3.1170274 \\
    6 & 1.7478474 & & 0.3956137 \\
    7 & 5.6966577 & & -0.3593987 \\
    8 & -0.5452435 & & -3.5266063 \\
    9 & 4.0826786 & & 2.4900307 \\
  \end{tabular}
\end{table}

We evaluated the performance of our synthesized DD sequence, UDD$_{12}$, and QDD$_{3}$ against the flip angle error $\Delta\beta = \varepsilon_1 = \varepsilon_3$ and the phase orthogonality error $\Delta\varphi$ = $\tan^{-1}(\varepsilon_2v_y(t)/v_x(t))$ on a single qubit system coupled to four bath TLSs, with the fidelity of quantum gate $\mathcal{C}$ defined as 
\begin{equation}
F({\mathcal C}) = 
 \text{min}_{|\phi\rangle}\| \sqrt{|\phi\rangle\langle\phi|}\sqrt{{\mathcal C}(|\phi\rangle\langle\phi|)}\|_{\text{tr}}. 
  \label{eq:gate_fidelity}
\end{equation}
The number of pulses and the pulse widths in UDD and QDD sequences were chosen such that the total amounts of the applied energy and the peak amplitudes were nearly equal to those of the generated waveform. If the energy of a DD sequence is defined as $\sqrt{\int_0^T \sum_{\alpha=x,y}v_\alpha^2(t)\,dt}$, the energies per cycle of the generated DD sequence, UDD$_{12}$, and QDD$_{3}$ amount to 11.50~$\pi/T$, 12~$\pi/T$, and 16~$\pi/T$, respectively. The peak amplitudes $\max_{t\in[0,T]}\sqrt{\sum_\alpha v^2_\alpha(t)}\, T/2\pi$ of the DD sequences were limited to 10, 20, and 22 for the synthesized DD waveform, UDD$_{12}$, and QDD$_{3}$, respectively. Given that there is minimal interpulse spacing in UDD and QDD, their peak amplitudes were not equal to that of the synthesized DD sequence. The coupling strengths between the qubit and TLSs were randomly generated; we used $\{g^{(k)}T/\pi\}_{k=1}^4$ = \{ 0.0338264, -0.0906347, 0.0014495, 0.0740895 \}, so that $\|\hat{\cal{H}}_0T\|/2\pi\sim0.1$.

As the flip angle errors are accumulated at each $\pi$ pulse, UDD is sensitive to the flip angle error (Fig.~\ref{fig:error}(a)); on the other hand, whereas QDD cancels the flip angle error by inserting a 90$^\circ$ phase-shifted $\pi$ pulse. However, as seen from Fig.~\ref{fig:error}(b), QDD cannot compensate for the phase orthogonality error. Figures~\ref{fig:error}(a) and (b) show that the generated sequence is robust against the pulse errors caused by the instruments. 
This example shows that a DD sequence can be optimized to be robust against known systematic errors, which are small but accumulate after millions of cycles in actual experiments. A variation of this example, where there is crosstalk between control pulses, can be managed in the same manner. 

Another variation is that the behavior of the surrounding bath TLSs is known, the Hamiltonian of which is denoted by $\hat{\mathcal{H}}_B$, e.g., the principal axes of TLSs are tilted from the $z$ axis so that $[\hat{\mathcal{H}}_{Q-B},\hat{\mathcal{H}}_B] \neq 0$. In this case, taking the bath Hamiltonian of the TLSs into account, we can obtain a DD sequence that overcomes degradation due to $\hat{\mathcal{H}}_B$ by minimizing higher-order cost functions $\Phi^{(n)}$, at least up to the second order. 

\section{Multiqubit Coupling\label{sec:mrev}}
Consider a system that consists of a finite number of qubits coupled to each other. Access to the qubits is assumed to be restricted to collective manipulation, the external perturbation of which is given by $\hat{V}(t) = v_x(t) \sum_k\hat{\sigma}_{x,k} + v_y(t) \sum_k\hat{\sigma}_{y,k}$. 
Each qubit is considered as being connected through a qubit-qubit Hamiltonian,
\begin{eqnarray}
  \hat{\mathcal{H}}_0 &=& 
  \hat{\mathcal{H}}_{Q-Q} = \sum_{k^\prime<k} d^{(k^\prime,k)} 
    \nonumber \\
  &\times&\left(\hat{\sigma}_{z,k^\prime}\hat{\sigma}_{z,k}
               -\hat{\sigma}_{x,k^\prime}\hat{\sigma}_{x,k}/2
               -\hat{\sigma}_{y,k^\prime}\hat{\sigma}_{y,k}/2
         \right).
  \label{eq:dipolarh}
\end{eqnarray}
In the interaction frame of $\hat{V}(t)$, we obtain a \textit{modulated} Hamiltonian,
\begin{equation}
\hat{\widetilde{\mathcal{H}}}_0(t) = \sum_{k^\prime,k}\sum_{\beta\gamma} d^{(k^\prime,k)} \eta_{\beta,\gamma}(t) \hat{\sigma}_{\beta,k^\prime}\hat{\sigma}_{\gamma,k},
\end{equation}
where $ \eta_{\beta\gamma}(t) = c_{\beta z}(t)c_{\gamma z}(t)  - c_{\beta x}(t)c_{\gamma x}(t)/2  - c_{\beta y}(t)c_{\gamma y}(t)/2$. We defined a cost function for the synthesis of $v_x(t)$ and $v_y(t)$ in a manner similar to that in Sec.~\ref{sec:dephase}, excluding $d^{(k^\prime,k)}$ from the cost function to manage the arbitrary strengths of the interactions and considering $\|\hat{\mathcal{H}}_{Q-Q}T\| < 1$. The synthesized waveforms expressed in a Fourier series are listed in Table~\ref{tab:ex2}.
\begin{figure}
 \centering
 \includegraphics{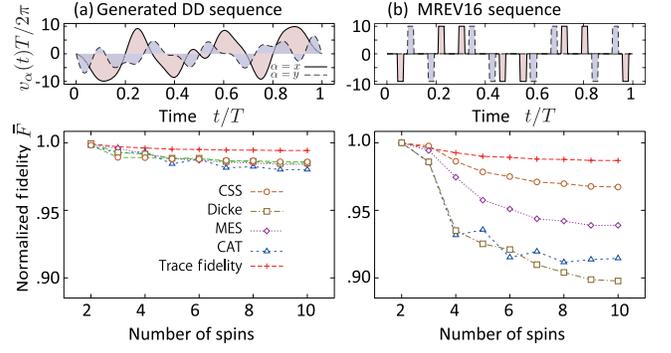}
 \caption{\label{fig:dipole}(color online). DD performance as evaluated for $n$-qubit chain system coupled through dipolar interactions with (a) optimized pulse sequence and (b) MREV16 sequence. Fidelity $F$ is normalized so that $\bar{F} = 1-(1-F)/\|\hat{\mathcal{H}}_{Q-Q}\|$ because $\|\hat{\mathcal{H}}_{Q-Q}\|$ becomes larger as the number of qubits increases.} 
\end{figure}

\begin{table}
  \caption{\label{tab:ex2}Parameters of the DD sequence in Sec.~\ref{sec:mrev}}
  \begin{tabular}{c|rlr}
    $n$ & $v_{x,n}T/\pi$ & \hspace{1em} & $v_{y,n}T/\pi$ \\
  \hline
    1 & -4.8892576 & & -2.6291726 \\
    2 & -3.1490576 & & -3.4112889 \\
    3 &-14.317448  & & -1.7326439 \\
    4 & -0.0929321 & &  4.2805093 \\
    5 &  6.8394959 & & -3.7925374 \\
    6 & -0.6645375 & & -2.3678092 \\
    7 &  0.3344480 & & -2.5797746 \\
    8 & -1.5042059 & & -1.9232075 \\
    9 &  2.3863574 & & -4.2795712 \\
  \end{tabular}
\end{table}

We evaluated the synthesized DD waveform by simulating a one-dimensional qubit chain by considering couplings up to the second nearest qubits, and compared its performance with that of MREV16~\cite{bib:Ladd05}, which is known to be powerful for this type of coupling. The Hamiltonian of the qubit chain was defined using the coupling strengths in Eq.~(\ref{eq:dipolarh}).
\begin{eqnarray}
  d^{(k,k^\prime)} = \left\{
  \begin{array}{lcl}
      \pi/T & & (k^\prime=k+1)\\
      \pi/8T & & (k^\prime=k+2)\\
      0 & & (\text{otherwise})\\
  \end{array}\right. .
\end{eqnarray}
Because it was difficult for us to calculate the gate fidelity defined in Eq.~(\ref{eq:gate_fidelity}) for several qubits, we used the state fidelity for typical quantum states and the trace fidelity defined as $F(\mathcal{C})=\text{Tr}[\sqrt{C^\dagger C}]$ instead, where $C$ is the unitary representation of $\mathcal{C}$. Four states were used and defined as follows: The coherent spin state directed toward the $x$-axis is the eigenstate of the collective $x$-operator $\sum_k \hat{\sigma}_{x,k}$ that has the maximum eigenvalue. The GHZ state is $|\text{GHZ}\rangle = |0\rangle^{\otimes n} + |1\rangle^{\otimes n}$, where $n$ is the number of qubits. The maximally entangled state is $|\text{MES}\rangle = 2^{-n/2} \sum_{i=0}^{2^n}|i\rangle$. The Dicke state is one of the eigenstates of the collective $z$-operator and the total spin operator $\hat{\mathbf{\sigma}}\cdot\hat{\mathbf{\sigma}}$ that have the $\hat{\sigma}_z$ eigenvalue closest to zero and the maximum $\hat{\mathbf{\sigma}}\cdot\hat{\mathbf{\sigma}}$ eigenvalue.
The energy and peak amplitude of this DD sequence were restricted to 13.8 $\pi/T$ and 20 $\pi/T$ so that they were almost equal to those of MREV16 (16 $\pi/T$ and 20 $\pi/T$, respectively).

Figure~\ref{fig:dipole} shows the fidelity per cycle against the number of the qubits in the system. The synthesized sequence effectively decouples qubit-qubit interactions for any state available in this example, whereas MREV16 shows considerable degradation in the nonclassical states owing to cooperative destruction caused by the finite width of the pulses and the qubit-qubit interactions. 
Given that the numerically synthesized sequence is optimized to isotropically suppress any operator coefficient of up to the first order, experimentally, it works better for the nonclassical states, which are necessary for QIP experiments.

\section{Performance versus resources\label{sec:resource}}
\begin{figure}
 \centering
 \includegraphics{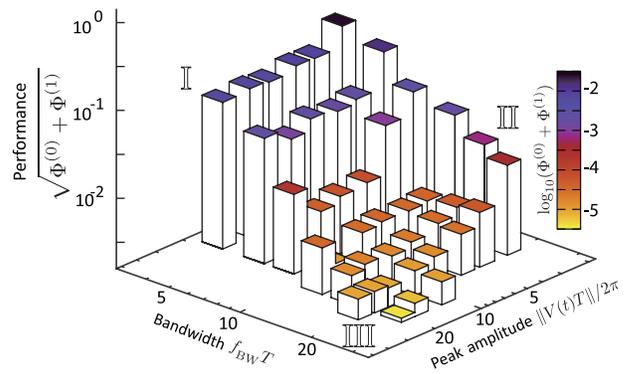}
 \caption{\label{fig:convergence}(color online). Performance under resource restrictions in terms of amplitude of $\|\hat{V}(t)T\|$ and bandwidth of $\hat{V}(t)$. Regions in which the maximum amplitude or the bandwidth is excessively supplied are labeled as I and II, respectively, and the region where both resources are sufficiently supplied is labeled as III. All axes are in log scale.}
\end{figure}

Our examples, with smoothly modulated wave DD sequences, are free from the finite pulse width problem and are robust against distortion in the waveforms under the finite frequency bandwidth limitation. Available resources can be specified in terms of the maximum control amplitude $\|\hat{V}(t)T\|$ and frequency bandwidth of $\hat{V}(t)$, so that DD sequence synthesis is performed within the available resources. We found that there is a close relationship among the DD performance, available bandwidth, and peak amplitude. Figure~\ref{fig:convergence} shows the performance of DD sequences used in Sec.~\ref{sec:mrev} under the resource restrictions. The root mean square of the operator coefficients in an average Hamiltonian up to the first order $\sqrt{\Phi^{(0)} + \Phi^{(1)}}$ was used as a measure of performance because this value indicates the suppression ratio of a coupling strength at $\|\hat{\mathcal{H}}_0T\| \sim 1$. We imposed frequency bandwidth limitation on $\hat{V}(t)$ using $p$ in Eq.~(\ref{eq:waveff}). Bandwidth of waveforms $f_{\text{BW}}$ was defined as $2p/T$.

The figure clearly shows that the performance of the generated DD sequences monotonically improves with an increase in the available resources; however, it also suggests that there is a trade-off between the performance and available resources. As seen from the figure, converged values of the cost function $\sqrt{\Phi^{(0)}+\Phi^{(1)}}$ are closely related to the available resources. This can also be seen in ref.~\cite{bib:Clausen10}, where the performance of synthesized DDs depends on available amplitude of $\hat{V}(t)$. Although the converged values are not always the global minima of $\Phi(\bm{\zeta})$ (because of the existence of local minima in nonlinear minimization), the figure suggests that available resources bound the performance of dynamical decoupling sequences. 

\begin{figure}
  \centering
  \includegraphics{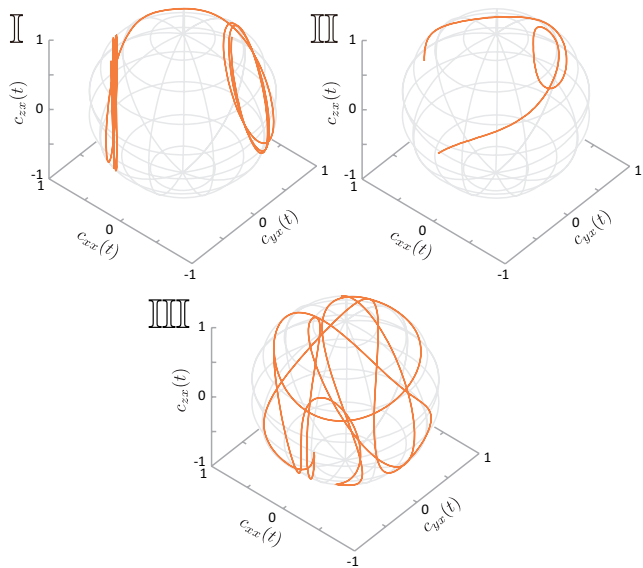}
  \caption{\label{fig:traj}(color online). System-modulation matrices $c_{\alpha x}(t)$ ($\alpha$ = $x, y, z$) in Regions I, II, and III, represented in Cartesian coordinates. $c_{\alpha x}(t)$ are displayed as functions of $t \in [0,T]$. }
\end{figure}

The figure shows another relationship between the resources: the decoupling efficiency drops when either of the resources is oversupplied. Region I in Fig.~\ref{fig:convergence} is ascribed to \textit{over driving}, which denotes lack of bandwidth of $\hat{V}(t)$ to utilize the large amount of driving energy for controlling the qubits. This behavior can be seen in Fig.~\ref{fig:traj}~(I), where a part of a system-modulation matrix $c_{\alpha x}(t)$ ($\alpha$ = $x$, $y$, $z$) in the region is represented as a trajectory in Cartesian coordinates. Given that the trajectory needs to have loops to consume the excess driving energy, the trajectory of the system-modulation matrix is constrained from moving efficiently to decouple, and consequently, the performance drops. Region II in Fig.~\ref{fig:convergence} is ascribed to \textit{over modulation} of $\hat{V}(t)$, wherein the amount of energy supplied for driving the qubits is insufficient~(Fig.~\ref{fig:traj}~(II)). When both resources are sufficiently supplied to drive and modulate, as in the region as III in Fig.~\ref{fig:convergence}, the system-modulation matrix in the region efficiently averages out unwanted interactions~(Fig.~\ref{fig:traj}~(III)).

\section{Conclusion}
We presented a general framework to numerically synthesize DD sequences, which can tailor them to a given set of qubits, environment, and realistic resource limitations. To this end, a cost function for the numerical synthesis of the DD sequences was built using system-modulation matrices of operators in Lie algebra, instead of a unitary propagator, in order to fully exploit the symmetries that exist in qubits and environments. The effects of finite pulse width are naturally incorporated in the system-modulation matrix. We presented two examples to demonstrate the robustness of the generated DD waveform to flip angle and phase orthogonality errors, regardless of the error amplitude, and the capability of the DD waveform eliminating unwanted qubit-qubit interactions, regardless of the number of qubits with protected nonclassical qubit states. The trade-off between performance and resource limitations was numerically studied. In the near future, we intend to rearrange a system Hamiltonian with the help of an external perturbation as introduced in \cite{bib:Brion11}, which is also applicable to QC and QIP experiments.

\begin{acknowledgments}
The authors thank M. Negoro for his fruitful discussions. This work was supported by the Funding Program for World-Leading Innovative R\&D on Science and Technology (FIRST), MEXT KAKENHI ``Quantum Cybernetics'' project. One of the authors (Y.T) was supported in part by the Japan Society for the Promotion of Science. 
\end{acknowledgments}

%\bibliography{tabuchi0}

\appendix
\section{Floquet Hamiltonian Theory\label{sec:floquet}}

A periodic time-dependent Hamiltonian $\hat{\widetilde{\mathcal{H}}}(t)$ can be averaged over a period $T$ using an approach based on the Floquet Theory. Consider a Fourier expansion of $\hat{\widetilde{\mathcal{H}}}(t)$ such that \begin{equation}
\hat{\widetilde{\mathcal{H}}}(t) = \sum_n\hat{\mathcal{H}}^{(n)}_{\text{F}} \exp\left(2i\pi nt/T\right). 
\end{equation}
As stated in \cite{bib:Scholz10}, average Hamiltonians up to second-order are given by 
\begin{eqnarray}
 \hat{\overline{\mathcal{H}}}_{\text{eff}}^{(0)} &=& 
  \hat{\mathcal{H}}^{(0)}_{\text{F}}, \label{eq:floquet_zeroth} \\
 \hat{\overline{\mathcal{H}}}_{\text{eff}}^{(1)} &=& 
  \frac{T}{4\pi} \sum_{n \neq 0} \frac{1}{n} 
   \left[\hat{\mathcal{H}}^{(n)}_{\text{F}},
      \hat{\mathcal{H}}^{(-n)}_{\text{F}}\right] 
            \label{eq:floquet_first}, \\
 \hat{\overline{\mathcal{H}}}_{\text{eff}}^{(2)} &=& 
  \frac{T^2}{12\pi^2}\sum_{n\neq 0}\sum_{n^\prime+n\neq 0} 
\frac{1+(1/2)\delta_{n^\prime,0}}{n(n+n^\prime)}
\nonumber \\ 
&\times&
  \left[\left[\hat{\mathcal{H}}^{(n)}_{\text{F}},
        \hat{\mathcal{H}}^{(n^\prime)}_{\text{F}}\right],
        \hat{\mathcal{H}}^{(-n-n^\prime)}_{\text{F}}
  \right].
\end{eqnarray}

\section{Derivation of average Hamiltonian for dephasing problem\label{sec:ex1_ah}}

The Hamiltonian of the system $\hat{\mathcal{H}}_0$ is considered as $\hat{\mathcal{H}}_0 = \hat{\mathcal{H}}_{Q-B} + \hat{\mathcal{H}}_{E}$ in the main text, where 
\begin{eqnarray}
 \hat{\mathcal{H}}_{Q-B} &=& \sum_k g^{(k)} \hat{\sigma}_z\hat{\sigma}_{z,k}, \\
 \hat{\mathcal{H}}_E &=& \varepsilon_1 v_x(t) \hat{\sigma}_x + \varepsilon_3 v_y(t) \hat{\sigma}_y \nonumber \\
&+& \varepsilon_2 v_y(t)\hat{\sigma}_x + \varepsilon_4 v_x(t)\hat{\sigma}_y,
\end{eqnarray}
and $\hat{\sigma}_\alpha$, $\hat{\sigma}_{\alpha,k}$ $(\alpha = x,y,z)$ denote Pauli operators of the qubit and the $k$th TLS, respectively.
A unitary operator $\hat{U}_V(t)$ = $\hat{\mathcal{T}}\exp(i\int_0^tV(t^\prime)dt^\prime)$ modulates the Hamiltonian $\hat{\mathcal{H}}_0$ in the interaction frame of $V(t)$ such that 
\begin{eqnarray}
 \hat{\widetilde{\mathcal{H}}}_{Q-B}(t) &=& \sum_{\alpha,k,n} g^{(k)} c_{\alpha z,n}\hat{\sigma}_{\alpha}\hat{\sigma}_{z,k} \,e^{2\pi int/T}, \\
   \hat{\widetilde{\mathcal{H}}}_E(t) &=& \sum_{\alpha,n} \left\{
     \varepsilon_1 \sum_{n^\prime}v_{x,n-n^\prime}c_{\alpha x,n^\prime} 
     \right. \nonumber \\
   &+& \varepsilon_2 \sum_{n^\prime}v_{y,n-n^\prime}c_{\alpha x,n^\prime} 
      \nonumber \\
    &+& 
     \varepsilon_3 \sum_{n^\prime}v_{y,n-n^\prime}c_{\alpha y,n^\prime}
      \nonumber \\
    &+& \left.\varepsilon_4 \sum_{n^\prime}v_{x,n-n^\prime}c_{\alpha y,n^\prime}
     \right\} \hat{\sigma}_\alpha \,e^{2\pi int/T}, 
\end{eqnarray}
where $v_{\alpha,n} = T^{-1}\int_0^T v_\alpha(t)\exp(-2\pi int/T)$ and $c_{\beta\alpha}(t) = \sum_n c_{\beta\alpha,n}\exp(2\pi int/T) = \text{Tr}[\hat{U}_V \hat{\sigma}_\alpha \hat{U}_V^\dagger \hat{\sigma}_\beta]/\text{Tr}[\hat{\sigma}_\beta^2]$. 
If we write 
\begin{eqnarray*}
d_{\alpha,n}^{(1)}=\sum_{n^\prime}v_{x,n-n^\prime}c_{\alpha x,n^\prime}, 
\hspace{.7em}
d_{\alpha,n}^{(2)}=\sum_{n^\prime}v_{y,n-n^\prime}c_{\alpha x,n^\prime}, 
  \\
d_{\alpha,n}^{(3)}=\sum_{n^\prime}v_{y,n-n^\prime}c_{\alpha y,n^\prime}, \hspace{.7em}
d_{\alpha,n}^{(4)}=\sum_{n^\prime}v_{x,n-n^\prime}c_{\alpha y,n^\prime},
\end{eqnarray*}
 the Fourier series of the \textit{modulated} Hamiltonians are 
\begin{eqnarray}
 \hat{\mathcal{H}}_{Q-B,\text{F}}^{(n)} &=& \sum_{\alpha,k} g^{(k)} c_{\alpha z,n}\hat{\sigma}_z\hat{\sigma}_{z,k}, \\
 \hat{\mathcal{H}}_{E,\text{F}}^{(n)} &=& \sum_{\alpha,i} \varepsilon_i \, d_{\alpha,n}^{(i)} \,\hat{\sigma}_\alpha .
\end{eqnarray}
The zeroth-order average Hamiltonian is easily obtained from $\hat{\overline{\mathcal{H}}}_{\text{eff}}^{(0)} = \hat{\mathcal{H}}_{Q-B,\text{F}}^{(0)} + \hat{\mathcal{H}}_{E,\text{F}}^{(0)}$, and is given by
\begin{equation}
 \hat{\overline{\mathcal{H}}}_{\text{eff}}^{(0)} = \sum_{\alpha,k} g^{(k)} c_{\alpha z,0}\hat{\sigma}_z\hat{\sigma}_{z,k} + \sum_{\alpha,i} \varepsilon_i \, d_{\alpha,0}^{(i)} \,\hat{\sigma}_\alpha .
 \label{eq:ave_zeroth}
\end{equation}
The first-order average Hamiltonian is obtained from Eq.~(\ref{eq:floquet_first}), and is given by 
\begin{eqnarray*}
 \hat{\overline{\mathcal{H}}}_{\text{eff}}^{(1)} &=& 
  \frac{T}{4\pi}\sum_{n\neq 0} \frac{1}{n}
    [\hat{\mathcal{H}}_{\text{F}}^{(n)},
    \hat{\mathcal{H}}_{\text{F}}^{(-n)}] \\
 &=& 
  \frac{T}{2\pi}\sum_{n > 0} \frac{1}{n}
    [\hat{\mathcal{H}}_{Q-B,\text{F}}^{(n)}+\hat{\mathcal{H}}_{E,\text{F}}^{(n)},
    \hat{\mathcal{H}}_{Q-B,\text{F}}^{(-n)}+\hat{\mathcal{H}}_{E,\text{F}}^{(-n)}] \\
 &=& 
  \frac{T}{2\pi}\sum_{n > 0} \frac{1}{n} \left\{
   [\hat{\mathcal{H}}_{Q-B,\text{F}}^{(n)},\hat{\mathcal{H}}_{Q-B,\text{F}}^{(-n)}]
  + [\hat{\mathcal{H}}_{E,\text{F}}^{(n)},\hat{\mathcal{H}}_{E,\text{F}}^{(-n)}]
  \right. \nonumber \\  
& & +\left.
    [\hat{\mathcal{H}}_{Q-B,\text{F}}^{(n)},\hat{\mathcal{H}}_{E,\text{F}}^{(-n)}]
  - [\hat{\mathcal{H}}_{Q-B,\text{F}}^{(-n)},\hat{\mathcal{H}}_{E,\text{F}}^{(n)}]
   \right\}, \nonumber 
\end{eqnarray*}
where
\begin{widetext}
\begin{eqnarray*}
\left[\hat{\mathcal{H}}_{Q-B,\text{F}}^{(n)},\hat{\mathcal{H}}_{Q-B,\text{F}}^{(-n)}\right] 
 &=& \sum_{k^\prime,k} g^{(k^\prime)}\,g^{(k)}
  \sum_{\alpha^\prime,\alpha} 
   c_{\alpha^\prime z,n} c_{\alpha z,-n}
  [\hat{\sigma}_{\alpha^\prime}\,,\hat{\sigma}_{\alpha}]
  \hat{\sigma}_{z,k^\prime}\,\hat{\sigma}_{z,k} \\
 &=& \sum_{k^\prime,k} g^{(k^\prime)}\,g^{(k)}
  \sum_{\alpha^\prime<\alpha} \left(
   c_{\alpha^\prime z,n} c_{\alpha z,-n} 
  - c_{\alpha z,n} c_{\alpha^\prime z,-n} 
                \right)
  [\hat{\sigma}_{\alpha^\prime}\,,\hat{\sigma}_{\alpha}]
  \hat{\sigma}_{z,k^\prime}\,\hat{\sigma}_{z,k} \\
 &=& \sum_{k^\prime,k} g^{(k^\prime)}\,g^{(k)}
  \sum_{\alpha^\prime<\alpha} \left(
   c_{\alpha^\prime z,n} c_{\alpha z,n}^*
  - c_{\alpha z,n} c_{\alpha^\prime z,n}^*
                \right)
  [\hat{\sigma}_{\alpha^\prime}\,,\hat{\sigma}_{\alpha}]
  \hat{\sigma}_{z,k^\prime}\,\hat{\sigma}_{z,k} \\
 &=& \sum_{k^\prime,k} g^{(k^\prime)}\,g^{(k)}
  \sum_{\alpha^\prime<\alpha} 
   2 \,\epsilon_{\alpha^\prime\alpha\beta} 
   \text{Im}[c_{\alpha z,n} c_{\alpha^\prime z,n}^*]\,
  \hat{\sigma}_{\beta}\,
  \hat{\sigma}_{z,k^\prime}\,\hat{\sigma}_{z,k}, \\
\left[\hat{\mathcal{H}}_{E,\text{F}}^{(n)},\hat{\mathcal{H}}_{E,\text{F}}^{(-n)}\right] 
 &=&
 \sum_{\alpha^\prime<\alpha}\sum_i \varepsilon_i^2\,
  2 \epsilon_{\alpha^\prime\alpha\beta}
  \text{Im}[d_{\alpha^\prime,n}^{(i)}\,d_{\alpha,n}^{(i)*}]\,
  \hat{\sigma}_\beta \\
 &+&\sum_{\alpha^\prime<\alpha}\sum_{i^\prime < i} 
  \varepsilon_{i^\prime} \varepsilon_{i}\,
  2 \epsilon_{\alpha^\prime\alpha\beta}
  \text{Im}[d_{\alpha^\prime,n}^{(i)}\,d_{\alpha,n}^{(i^\prime)*}
       + d_{\alpha^\prime,n}^{(i)*}\,d_{\alpha,n}^{(i^\prime)}]\,
  \hat{\sigma}_\beta, \\
 \left[\hat{\mathcal{H}}_{Q-B,\text{F}}^{(n)},\hat{\mathcal{H}}_{E,\text{F}}^{(-n)}\right]
 &=& \sum_k \sum_i g^{(k)} \varepsilon_i \sum_{\alpha^\prime < \alpha}
  2 \epsilon_{\alpha^\prime\alpha\beta}
  \text{Im}[c_{\alpha^\prime z,n}^*\,d_{\alpha,n}^{(i)}
       + c_{\alpha z,n}\,d_{\alpha^\prime,n}^{(i)*}]\,
   \hat{\sigma}_\beta\,\hat{\sigma}_{z,k},
\end{eqnarray*}
\end{widetext}
and $\epsilon_{\alpha\beta\gamma}$ is the Levi-Civita symbol. 
$\sum_{\alpha^\prime < \alpha}$ represents the summation of $(\alpha^\prime,\alpha)=\{(x,y),(x,z),(y,z)\}$. 
Here, we used the identity $c_{\alpha\beta,-n} = c_{\alpha\beta,n}^*$ because $c_{\alpha\beta}(t)$ is real. The first-order average Hamiltonian then becomes
\begin{widetext}
\begin{eqnarray}
 \hat{\overline{\mathcal{H}}}_{\text{eff}}^{(1)} 
&=& 
 \frac{T}{\pi}
 \sum_{k^\prime,k} g^{(k^\prime)}\,g^{(k)}
  \sum_{\alpha^\prime<\alpha} \,
  \varepsilon_{\alpha^\prime\alpha\beta} 
  \left(
  \sum_{n > 0} \frac{1}{n}
   \text{Im}[c_{\alpha z,n} c_{\alpha^\prime z,n}^*]
  \right)\,
  \hat{\sigma}_{\beta}\,
  \hat{\sigma}_{z,k^\prime}\,\hat{\sigma}_{z,k} \nonumber \\
&+&
 \frac{T}{\pi}
 \sum_{\alpha^\prime<\alpha}\sum_i \varepsilon_i^2\,
  \epsilon_{\alpha^\prime\alpha\beta}
 \left(
 \sum_{n>0} \frac{1}{n}
  \text{Im}[d_{\alpha^\prime,n}^{(i)}\,d_{\alpha,n}^{(i)*}]
 \right)\,
 \hat{\sigma}_\beta \nonumber \\
&+&
 \frac{T}{\pi}
 \sum_{\alpha^\prime<\alpha}\sum_{i^\prime < i} 
  \varepsilon_{i^\prime} \varepsilon_{i}\,
  \epsilon_{\alpha^\prime\alpha\beta}
 \left(
 \sum_{n>0} \frac{1}{n}
  \text{Im}[d_{\alpha^\prime,n}^{(i)}\,d_{\alpha,n}^{(i^\prime)*}
       + d_{\alpha^\prime,n}^{(i)*}\,d_{\alpha,n}^{(i^\prime)}]
  \right)\,\hat{\sigma}_\beta \nonumber\\
 &+& 
 \frac{2T}{\pi}
  \sum_k \sum_i g^{(k)} \varepsilon_i \sum_{\alpha^\prime < \alpha}
  \epsilon_{\alpha^\prime\alpha\beta}
 \left(
 \sum_{n>0} \frac{1}{n}
  \text{Im}[c_{\alpha^\prime z,n}^*\,d_{\alpha,n}^{(i)}
       + c_{\alpha z,n}\,d_{\alpha^\prime,n}^{(i)*}]
 \right)\,\hat{\sigma}_\beta\,\hat{\sigma}_{z,k}.
 \label{eq:ave_first}
\end{eqnarray}
\end{widetext}

\section{Derivation of the cost function for dephasing problem\label{sec:ex1_cost}}
The zeroth-order cost function $\Phi^{(0)}$ is defined from Eq.~(\ref{eq:ave_zeroth}) as follows:
\begin{equation}
 \Phi^{(0)} = \sum_{\alpha = x,y,z}
    \left\{
    \left|c_{\alpha z,0}\right|^2 
    + w^2 \sum_i \left|d_{\alpha,0}^{(i)}\right|^2
    \right\}.
\end{equation} 
The first order cost function $\Phi^{(1)}$ is defined such that terms within the curly bracket in Eq.~(\ref{eq:ave_first}) become zero, and is given by
\begin{eqnarray}
\Phi^{(1)}
&=&
  \sum_{\alpha^\prime<\alpha} \,
  \left|
  \sum_{n > 0} \frac{1}{n}
   \text{Im}[c_{\alpha z,n} c_{\alpha^\prime z,n}^*]
  \right|^2 \nonumber \\
& & + w^4
 \sum_{\alpha^\prime<\alpha}\sum_i
 \left|
 \sum_{n>0} \frac{1}{n}
  \text{Im}[d_{\alpha^\prime,n}^{(i)}\,d_{\alpha,n}^{(i)*}]
 \right|^2 \nonumber \\
&+& w^4
 \sum_{\alpha^\prime<\alpha}\sum_{i^\prime < i} 
 \left|
 \sum_{n>0} \frac{1}{n}
  \text{Im}[d_{\alpha^\prime,n}^{(i)}\,d_{\alpha,n}^{(i^\prime)*}
       + d_{\alpha^\prime,n}^{(i)*}\,d_{\alpha,n}^{(i^\prime)}]
 \right|^2 \nonumber \\
&+& w^2
  \sum_{\alpha^\prime < \alpha} \sum_i 
  \left|
  \sum_{n>0} \frac{1}{n}
  \text{Im}[c_{\alpha^\prime z,n}^*\,d_{\alpha,n}^{(i)}
       + c_{\alpha z,n}\,d_{\alpha^\prime,n}^{(i)*}]
  \right|^2, \nonumber \\
  & &
\end{eqnarray}
where $w$ is the deliberate weight for the pulse errors set to 1/100.

\end{document}